\documentclass[twocolumn,prl,amsmath,amssymb,showkeys,showpacs]{revtex4}
\usepackage{graphicx}
\usepackage{bm}

\begin{document}

\title{\bf  Light scalars coupled to photons and non-newtonian forces}

\author{Arnaud Dupays}
\affiliation{INAF-IASF, Via E. Bassini 15, I-20133 Milano,
  Italy
}
\author{Eduard Mass{\'o}}
\affiliation{Grup de F{\'\i}sica Te{\`o}rica and Institut de F{\'\i}sica
  d'Altes Energies, Universitat Aut{\`o}noma de Barcelona, 08193 Bellaterra,
  Spain}

\author{Javier Redondo}
\affiliation{Grup de F{\'\i}sica Te{\`o}rica and Institut de F{\'\i}sica
  d'Altes Energies, Universitat Aut{\`o}noma de Barcelona, 08193 Bellaterra,
  Spain}
\author{Carlo Rizzo}
\affiliation{Laboratoire Collision Agr{\'e}gats et R{\'e}activit{\'e}, IRSAMC,
  Universit{\'e} Paul Sabatier et CNRS, 31062 Toulouse, France}








\date{Preprint UAB-FT-618}

\begin{abstract}
A particle $\phi$ coupling to two photons couples also radiatively to charged particles, like protons.
If the particle is a light scalar this induced coupling leads to spin-independent non-newtonian forces.
We show that the experimental constraints on exotic, fifth-type forces lead to stringent
constraints on the $\phi\gamma\gamma$ coupling. We discuss the impact on the recent PVLAS results
and the role of paraphoton models introduced to solve the PVLAS-CAST puzzle.
\end{abstract}

\maketitle

Spinless light particles are a common prediction of many theories that go beyond the
standard $SU(3)_c\times SU(2)_L \times U(1)_Y$ gauge theory. Probably the most famous
of them is the axion \cite{WeinbergEM}, an unavoidable consequence of the introduction
of a new global $U(1)$ symmetry designed to solve the QCD CP-problem \cite{Peccei2EM}. There are other
examples of light particles; some are pseudoscalar like the axion itself and some are scalar particles. We call them axionlike particles (ALPs).

In general, a spinless particle couples to two photons, and of course in principle
it has also couplings to matter. But the $\gamma\gamma$ coupling is particularly
interesting because many experimental searches for ALPs are based on it. One of these searches
is based on the so-called haloscope \cite{SikivieEM}, where axions or some other similar particles forming
part of the  dark matter in the galactic halo can convert into photons in a cavity with a strong magnetic
field. Another search is to look for ALPs produced in the Sun
converting into photons in a detector having a strong magnetic field; this is called helioscope \cite{SikivieEM}. Still
another experiment searching for ALPs and that uses the $\gamma\gamma$ coupling
looks for optical dichroism and birrefrigency in a laser that propagates in a magnetic field
\cite{MaianiEM}.

While the most recent haloscope \cite{haloscope} and helioscope \cite{ZioutasEM} experiments
have not found any signal and thus have put limits on model parameters, the third type of
experiment described above has reported a positive signal. Indeed, the PVLAS collaboration
finds a rotation of the polarization plane of the laser as well as an induced
ellipticity \cite{ZavattiniEM}. There is an exciting interpretation of these results in
terms of a new scalar particle $\phi$, that should have a mass $m_\phi \sim 10^{-3}$ eV and
a $\phi\gamma\gamma$ coupling scale of about $M \sim 10^5$ GeV -see equation
(\ref{Lone}) below for the definition of $M$.

The purpose of this paper is to show that in the case that the particle $\phi$ is indeed
a light {\bf scalar}, the $\phi\gamma\gamma$ coupling leads to the existence of long-range spin-independent
non-newtonian forces. Our calculation will allow us to find a very stringent limit
on the coupling, using the experimental null results coming from searches for new forces.

A scalar particle couples to two photons with the lagrangian
\begin{equation}
{\cal L}_1= {1 \over 4 M}\, \phi F^{\mu\nu} F_{\mu\nu}
\label{Lone}
\end{equation}
The key point is that (\ref{Lone}) induces radiatively a coupling to charged particles,
for example to protons (see Fig 1). The induced coupling will have the form
\begin{equation}
{\cal L}_2= y\, \phi \bar \Psi \Psi
\label{Ltwo}
\end{equation}
where $\Psi$ is the proton field and $y$ the Yukawa coupling. The loop diagram  is logarithmically divergent.

In order to treat physically this divergence, we notice that (\ref{Lone})
corresponds to a non-renormalizable term in the lagrangian and as such is expected
to be valid only up to a high-energy scale $\Lambda$, where new physics has to appear.
We integrate momenta in the loop of Fig.1 only until $\Lambda$, so that we cut
the divergence with the scale $\Lambda$.

The logarithmically divergent part of the diagram of Fig.1 is well-defined and is the leading radiative contribution to $y$. Approximating $y$ by this term we obtain
\begin{equation}
y= {3 \over 2} \, {\alpha \over \pi}\,  {m_p \over M}\,  \log {\Lambda \over m_p}
\label{y}
\end{equation}

In principle there is also a tree level Yukawa term (\ref{Ltwo}) in the theory, and there could be some cancelation between the tree level and the radiatively induced coupling
(\ref{y}). We regard this possibility as very unnatural, and will use (\ref{y}) for our estimates.

\begin{figure}
  \includegraphics[width=5cm]{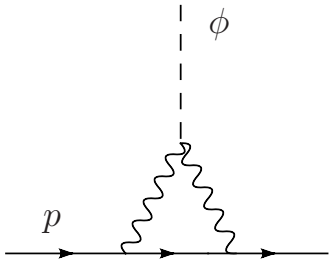}\\
  \caption{Loop diagram.}
\end{figure}

The Yukawa coupling (\ref{Ltwo}) leads to non-newtonian forces between two test masses $m_1$ and $m_2$,
due to $\phi$ exchange. The total potential between 1 and 2 is
\begin{equation}
V(r) = G {m_1 m_2 \over r}\, + \, \frac{y^2}{4\pi} {n_1 n_2 \over r} e^{- m_\phi r}
\label{V}
\end{equation}
Here $n_i$ is the total number of protons in the test body $i$. In (\ref{V}) we have neglected
the new electron-proton and electron-electron force since the corresponding value for $y$ in
the case of the electron is smaller than (\ref{y}) by a factor $m_e/m_p$.
When the two test bodies are constituted each by only one element, with atomic numbers $Z_1$
and $Z_2$, and mass numbers $A_1$ and $A_2$, we can approximate (\ref{V}) by
{\small
\begin{equation}
V(r) \simeq G {m_1 m_2 \over r}\, \left[ 1 + {1 \over G m_p^2} \, \frac{y^2}{4\pi} \,
\left( {Z \over A} \right)_1 \left( {Z \over A} \right)_2 e^{- m_\phi r} \right]
\label{Vapprox}
\end{equation}
}where now in the second term inside the square brackets i.e., the term containing the
correction to the newtonian potential, we have approximated $m_i\simeq A_i m_p$.

The non-newtonian part of the potential we have obtained has two
clear properties: it has a finite range $m_\phi^{-1}$ and depends on the composition of the bodies, i.e. on their $Z/A$ values.
We can use the abundant experimental bounds on fifth-type forces to limit our parameter $M$ as a function of $m_\phi$.

To see how we proceed we find now the limit in the interesting ranges
$m_\phi^{-1} \sim ({\rm meV})^{-1} \sim 0.2$ mm, with the PVLAS results in mind. Bounds on new forces have been obtained by experiments designed to
measure very small forces. In the submillimeter range, in 1997
authors of ref. \cite{carugnoetal} using a micromechanical
resonator designed to measure Casimir force between parallel
plates gave limits on new forces, and they also estimated the
corresponding limits on the mass and inverse coupling constant of
scalar particles. Also, strong experimental limits on
new forces have been published in \cite{Hoyleetal},  where they
 use a torsion pendulum and a rotating attractor in
the framework of tests of the gravitational inverse-square law.
The most strict bounds have been obtained very recently by using
torsion-balance experiments \cite{kapner}.

The limit from the experiment presented in ref.
\cite{kapner} for $m_\phi = 10^{-3}$ eV is
\begin{equation}
\frac{y^2}{4\pi}  {1 \over G m_p^2} \left( {Z \over A} \right)_1 \left( {Z \over
A} \right)_2 < 1.3 \times 10^{-2}
\end{equation}
which leads, introducing the conservative values for $(Z/A)$ of
0.4,
\begin{equation}
y < 7.8 \times 10^{-20} \label{y_bound}
\end{equation}

To obtain now a limit on $M$, we will simply put the value of the
log in (\ref{y}) equal to 1; this will lead to a conservative
limit since $\Lambda \gg m_p$. With this, we obtain
\begin{equation}
M >  4.2 \times 10^{16}\ {\rm GeV} \label{M_bound}
\end{equation}

Such a high lower bound implies that no signal of a $0^+$ particle of mass
$m_\phi \sim $ meV should be seen in experiments like PVLAS \cite{ZavattiniEM} or CAST
\cite{ZioutasEM}. In order to reach our conclusion we have to assume that the lagrangian
(\ref{Lone})  is valid up to high energy scales $\Lambda \gg m_p$.

\begin{figure}
   \includegraphics[width=8cm]{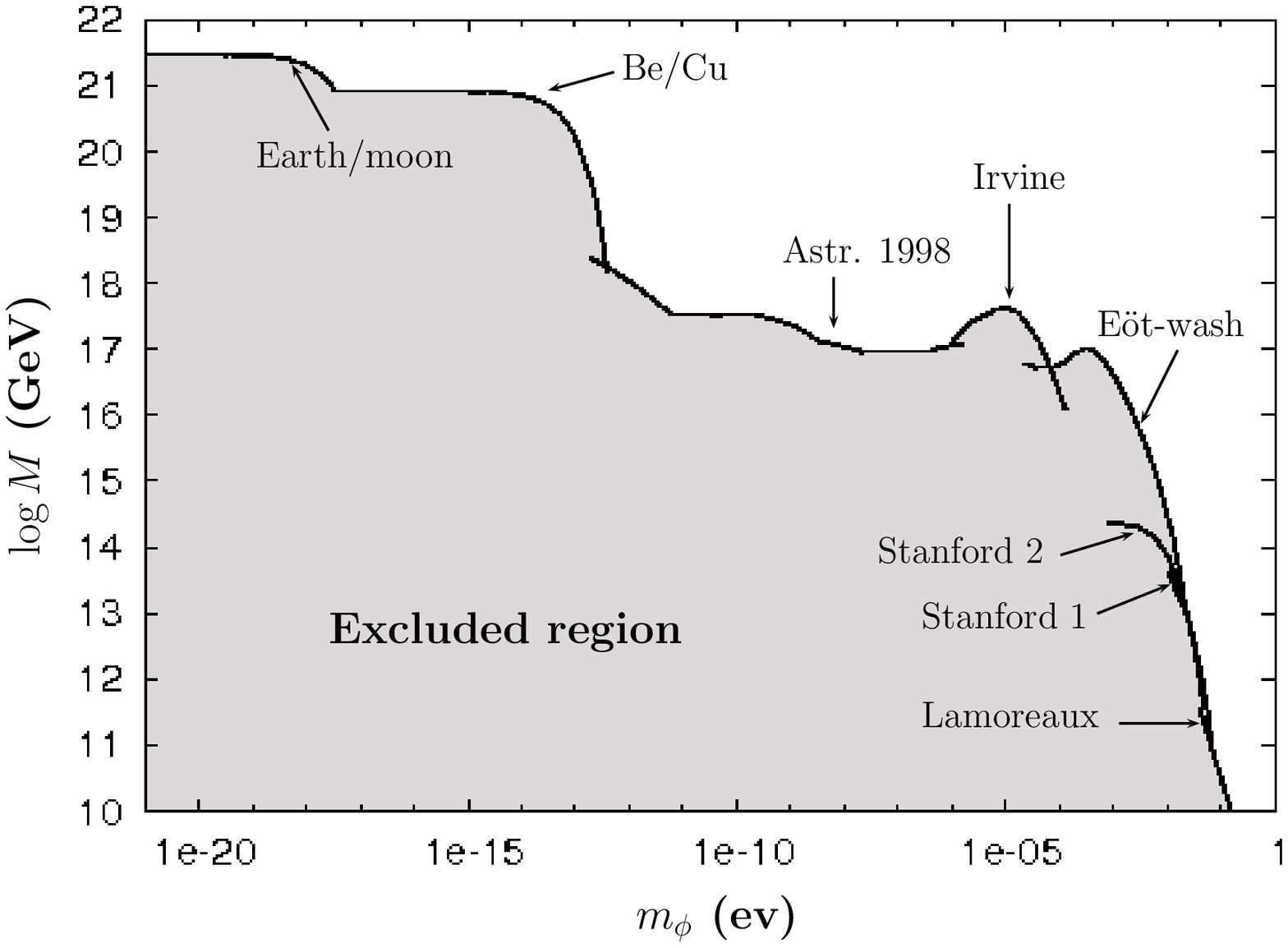}\\
  \caption{Constraints in the $\log{M}-m_\phi$ plane. Lines labeled Earth/moon
  and Astr. 1998 show constraints from astrophysical observations,
  Refs. \cite{EarthMoon} and \cite{Onofrio} respectively. Lines labeled Be/Cu,
  Irvine, E\"ot-wash, Stanford 2, Stanford 2, and Lamoreaux show experimental
  constraints, Refs. \cite{BeCu}, \cite{Irvine}, \cite{kapner}, \cite{Stanford:2},
  \cite{Stanford:1}, and \cite{Lamoreaux} respectively. The shaded region is excluded.}
\end{figure}

We can proceed in analogous way and find bounds for other values of
$m_\phi^{-1}$. They  are shown in Fig. 2, and as we see, the limits on $M$ are very tight.

Our final discussion is about modified   $\phi\gamma\gamma$ vertices.
Examples of such models have been developed in \cite{Masso:2005ym} with the motivation of
making compatible the PVLAS particle interpretation with the bounds coming from stellar
energy loss and a fortiori with the CAST results. In these models the ALP does couple to new paraphotons and has not a direct coupling to photons.
The   $\phi\gamma\gamma$ vertex arises because there is kinetic  photon-paraphoton
mixing. Also, a paraphoton mass $\mu$
induces an effective photon form factor such that the coupling is reduced for $|q|\gg \mu$.

When there is a
form factor with a low scale $\mu$, the analysis shown in this paper should be modified accordingly. In such a
case the lagrangian in (\ref{Lone}) is valid only for photons with momentum $q$ such that  $|q| \ll \mu$.
The induced
Yukawa will be suppressed with respect to the value (\ref{y}), and thus
the lower bound (\ref{M_bound}) can be very much relaxed if indeed $\mu$ is a low energy scale.

In order to quantify our last assertion, we have calculated the induced Yukawa coupling when the photons have a modified propagator
\begin{equation}
{1\over q^2} \rightarrow {1 \over q^2} \,  {\mu^2 \over \mu^2 - q^2}
\label{ff}
\end{equation}
Now the diagram of Fig.1 is finite.  The calculation of the coupling to protons  $y'$,
 at leading order in $\mu/m_p$ gives
\begin{equation}
y'=  {\alpha \over 4}\,  {\mu \over M}
\label{yp}
\end{equation}

To find the potential between bodies
we have to take into account the coupling to protons as well as to electrons,
because (\ref{yp}) is independent of the mass of the fermion.
The potential when having a form factor (\ref{ff}) is given by
{\small
\begin{equation}
V(r) \simeq G {m_1 m_2 \over r}\, \left[ 1 + {4 \over G m_p^2} \, \frac{y'^2}{4\pi} \,
\left( {Z \over A} \right)_1 \left( {Z \over A} \right)_2 e^{- m_\phi r} \right]
\label{Vff}
\end{equation}
}
We see that the new non-standard potential (\ref{Vff}) has a Yukawa coupling $y'$ that compared  to (\ref{y})  is suppressed  by a factor of order $\mu/m_p$. The parameter
$\mu$ introduced in  \cite{Masso:2005ym} is not fully specified by the theory. In order to solve the PVLAS-CAST puzzle and if we do not wish too different scales in the model, $\mu$ should be in the subeV range with a preferred value $\mu \sim $ meV. For this value, the bound (\ref{M_bound}) would
relax by 12 orders of magnitude, bringing it close to the PVLAS value $M \sim 10^5$ GeV.
Remarkably enough, experiments searching for new forces and testing Casimir forces at the submm
lengths may be sensitive to the potential (\ref{Vff}) that corresponds to a vertex with the
form factor (\ref{ff}).

{\bf Note added :} The fact that a scalar-photon-photon coupling gives rise to new forces
and leads to a bound on $M$ has been independently realized by Shmuel Nussinov
\cite{zioutas}.

\section*{Acknowledgments} EM and JR acknowledge support by the CICYT Project
FPA2005-05904 and the DURSI Project 2005SGR00916. We thank
Pierre Sikivie for pointing out to us that a form factor relaxes the bound on $M$,
and Holger Gies and Seth Hoedl for useful discussions.

\end{document}